# "Cracking-the-whip" effect stretches driven polymers

Juan Li,[1,2] Wenbing Hu[1,3,*]

[1]State Key Laboratory of Coordination Chemistry, School of Chemistry and Chemical Engineering, Nanjing University, 210093 Nanjing, China

[2]School of Mechanical Engineering, Nanjing Institute of Industry Technology, 210023 Nanjing, China

[3]Kavli Institute for Theoretical Physics China at the Chinese Academy of Sciences, 100190 Beijing, China.

*E-mail address: wbhu@nju.edu.cn

**Cracking the whip accelerates the tail of a chain to hit the air loudly and clearly. We proved that the similar acceleration effect causes coil deformation of driven chain-like polymers. We first preformed Monte Carlo simulations of a single driven polymer coil to demonstrate its deformation in company with faster or slower deviations of velocities. We then performed parallel Brownian Dynamics simulations to demonstrate that the coil deformation can be caused by the so-called "cracking-the-whip" effect due to non-synchronous biased Brownian motions of monomers inherited in Monte Carlo simulations. Since such non-synchronous motions represent random perturbations in the environmentally dependent potential energy landscape or mobility, reflecting heterogeneous dynamics of polymers in the liquid phase, our observations bring new insights into the non-linear dynamics of driven chain-like polymers.**

## I. INTRODUCTION

The lattice Ising model of bulk atomic and molecular particles has been well established as an equilibrium model for the study of their phase transitions.[1-5] For lattice particles performing Brownian motions, Monte Carlo (MC) simulations of phase transitions commonly employ Metropolis sampling algorithm on the basis of the detailed balance[6,7] that is equivalent to Onsager's reciprocal relation with a linear approximation near equilibrium.[8] Therefore, this algorithm is legitimate in the study of non-equilibrium evolutions of Brownian particle systems, such as kinetics and morphology of phase transitions. Under a situation far away from equilibrium, for instance, under an external force field, the detailed balance will be broken down; however, the system evolution can still be informative along this approach, provided that the Brownian motions of particles are modeled reasonably well. In 1983, by means of MC simulations of driven lattice gas to bias the originally random hopping of molecules into its neighboring vacancy sites along a specific direction, Katz, Lebowitz and Spohn observed a spontaneous density heterogeneity along the vertical direction,[9] which has attracted a broad attention in the field of non-equilibrium statistical mechanics.[10]

What if all the driven Brownian particles are constrained in a polymer chain? This question appears practically meaningful to those bulk polymers in the driven flows where hydrodynamic interactions between polymers can be effectively screened off by dense interpenetration of polymer coils.[11] Furthermore, answering this question will facilitate our better understanding to those non-linear dynamics (and thus non-linear viscoelastic behaviors) of driven polymers in many practically important cases, such as interfacial migration, centrifugal/gravitation precipitation, electrophoresis, and driven flows upon polymer processing. So far, little attention along this approach has been paid to the driven polymers in comparison to the driven lattice gas.

Amorphous polymer chains intend to be random coils due to their diffusion integrated by the Brownian motions of monomers.[11] Therefore, the quiescent dynamics of bulk long-chain polymers has been described as the sliding diffusion of Rouse-chain polymers along the primitive path of random coil.[12,13] Polymer diffusion driven by the external fields is also integrated by the biased Brownian motions of monomers. In principle, an ideal integration of biased Brownian motions of monomers performs only parallel transport in a constant force field, and brings no deformation to the polymer coil. But the reality could be far more complicated than this ideal situation.

In this paper, by means of dynamic MC simulations of a single phantom free-draining bead-stick lattice polymer in a constant force field, we observed a spontaneous coil deformation raised by strong driving forces. Since MC simulations underwent non-synchronous activations randomly distributed along a single chain, we performed Brownian Dynamics (BD) simulations of a single phantom free-draining bead-spring polymer to demonstrate that non-synchronous activation could be responsible for the observed coil-deformation upon biased Brownian motions of monomers. This deformation mechanism revealed an acceleration effect along the chain similar to the case of "cracking-the-whip", which appears fundamental to the non-linear viscoelastic behaviors of driven polymers.

## II. TECHNIQUES OF MONTE CARLO SIMULAITONS

We performed dynamic MC simulations of a bead-stick lattice polymer model to study a free-jointed phantom single chain in a constant force field imposed onto all the monomers. The single chain here represented those polymers in the bulk phase where the hydrodynamic interactions have been screened off by dense interpenetration of polymer coils; therefore, no hydrodynamic interaction beyond each monomer was considered. The single chain occupies consecutive lattice sites in a $64^3$ cubic lattice box with periodic boundary conditions, and the empty single sites stand for the phantom theta solvent surrounding the polymer. Double occupation and bond crossing were allowed on mimicking an unperturbed chain in the bulk phase of polymers. In each trial move of micro-relaxation, the randomly selected monomer tried to hop into its vacancy neighbor if available, sometimes accompanied with local sliding diffusion along the chain to maintain bond lengths within the unit of lattice

spacing.[14,15]

In each trial move of micro-relaxation, we employed a constant force field to make monomers move preferably along Y-axis. The reduced driving-force potential $F$ was given by

$$F = \frac{fu}{k_B T},$$  (1)

where $f$ was the strength of external forces conducted presumably by the momentum, which was transferred by the surroundings of each monomer (collisions exempting the hydrodynamic drag forces), $u$ was the unit of lattice spacing, $k_B$ was the Boltzmann's constant and $T$ the temperature. The potential along Y-axis was imposed onto each monomer by the well-known Metropolis importance sampling. Such an algorithm of MC simulations has been applied in the study of driven lattice gas[9] as well as of polymer driven flows.[16-22] In details, the net number of forward jumps, $n$, was calculated in each trial move of micro-relaxation. If $n$ is positive, the move will be accepted; while if $n$ is negative, its acceptability depends on the Boltzmann's exponential of $nF$.

The initial state of the single chain for biased diffusion in our simulations was a random coil prepared via a long-term relaxation process. The relaxation of the chain began from a hairpin conformation oriented along Y-axis. The unit of the time period was defined as one MC cycle, which was the number of trial moves equivalent to the total number of monomers. Here, random selection of monomers to hop into its vacant neighbor provides a high efficiency of sampling acceptance, which decides the speed of a MC procedure converging to the Brownian dynamics of small particles, as evidenced by a comparison with the parallel BD simulations.[23] Mean square radius of gyration $R^2$ was traced to represent the coil sizes according to

$$R^2 = < \sum_{j=1}^{N} (r_j - r_{cm})^2 \Big/ N >$$  (2)

where $r_j$ was the coordinate of the $j$-th monomer and $r_{cm}$ was the coordinate of the coil mass center. When three dimensional fractions of coil sizes meet together, the polymer has approached at the random-coil conformation. The coordinates of the monomers at the end of the relaxation period were then recorded as the initial state of the driven polymer. Since the hydrodynamic interactions have been exempted, present simulations access only to the Rouse-chain dynamics in the free-draining limit of short chains in the bulk polymer phase.[24] To identify the reasonably well Brownian motions of short chains in our simulations, we calculated the mean-square displacement of monomers $g(t)$ according to

$$g(t) = < r(t) - r(0) >^2$$  (3)

where $r(t)$ was the position of a monomer at time $t$ and $<...>$ meant an average over all the monomers. Figure 1 demonstrates the dynamic scaling of single chains switching from the self-diffusion to the biased diffusion with a constant velocity. The curve without driving force changes its slopes from 0.5 to 1, following the scaling laws of Rouse chains when diffusing out of their contour sizes. The curve with driving forces changes its slopes from 1 to 2, indicating a constant activated velocity.

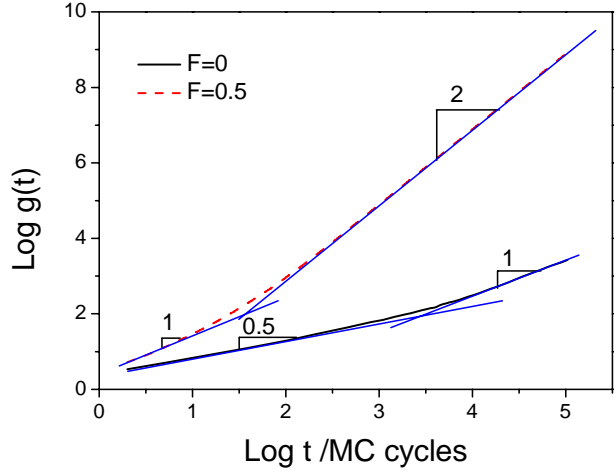

**Figure 1** Log-log plots of mean square displacements $g(t)$ of monomers for a single phantom chain containing 128 monomers versus evolution time $t$ with and without driving forces as denoted. Only the curves are drawn without showing the data points for clarity, and each curve has been averaged over 100 repeating times. The straight lines are drawn with their dominated slopes to guide the eyes.

## III. RESULTS OF MONTE CARLO SIMULATIONS

We first raised the driving forces step-by-step from zero to ten, and then brought them back in the same steps from ten to zero. Long-term relaxation was given at each step, and the deformation of the polymer coil was monitored by the departure between the parallel ($R_Y^2$) and perpendicular ($(R_X^2 + R_Z^2)/2$) fractions of mean square radius of gyration. Such a departure in the fractions of coil sizes reflects the first normal-stress difference often used in the characterization of non-Newtonian-fluid behaviors of flowing polymers. Figure 2a demonstrates clearly the coil deformation of the single chain upon the enhancement of driving forces. The exactly reversible path upon the weakening of driving forces implies the steady states of deformation at each step. The normal-stress differences under large driving forces appear as saturated at an up-limit of conformational-entropy elasticity, although the saturated sizes are still far from the fully-stretched size of the chain. This deformation behavior is consistent with our recent dynamic MC simulations of bulk lattice polymer chains driven in a constant force field,[21] implying its single-chain origin.

Visual inspections on the spatial distributions of monomers around the mass center of the chain demonstrate more straightforward the deformation, as shown in Fig. 2b. One can see that, with the enhancement of driving forces, the chain has been deformed from a coil to a stretched state. The stretching of the chain appears quite asymmetric with a relatively large head and a small tail along the force field. More asymmetric deformation on the shapes of a single coil has been observed for the single tethered chain, with one chain end pinned and the other end subject to a uniform flow.[25-27]

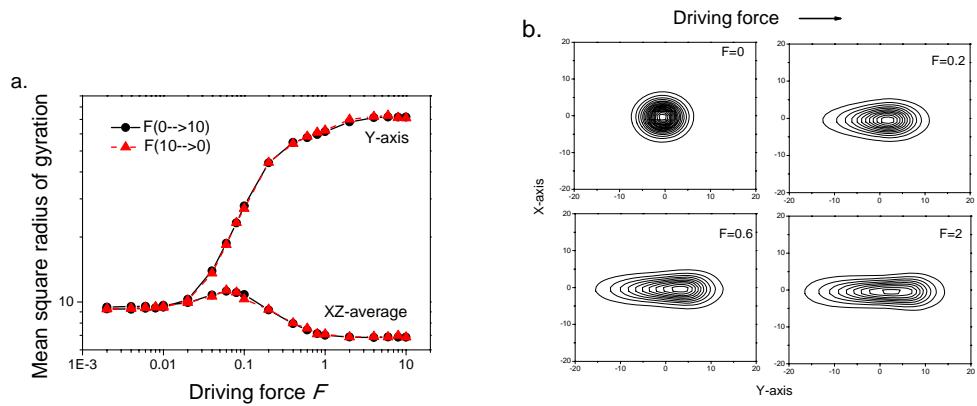

**Figure 2 (a)** Parallel fraction (upside) and perpendicular fraction (downside, averaged over two sectional dimensions) of mean square radius of gyration for the single phantom chain containing 128 monomers upon step-by-step increasing (spheres) and then decreasing (triangles) of the driving forces in our Monte Carlo simulations. **(b)** Contour plots of two-dimensional projections of monomer densities around the mass center of the single coil under various strengths of driving forces ($F$). The data was collected as an ensemble average over $10^8$ MC cycles at each step.

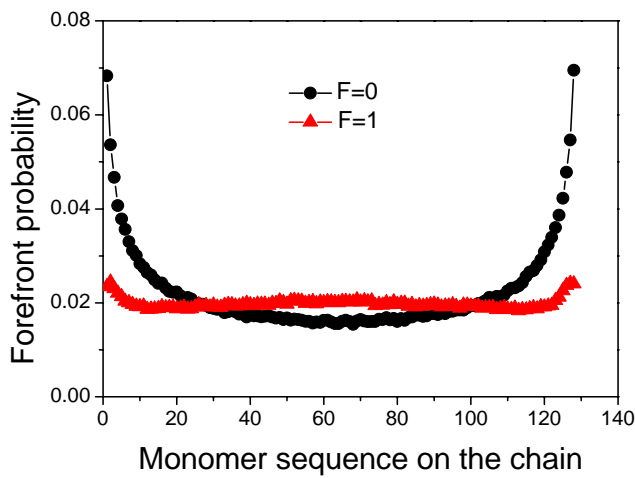

**Figure 3** Probability distributions of monomer sequences (from one to 128) occurring at the diffusion forefront of a single coil. Different driving forces ($F$=0 and 1) were employed onto the Y-axis direction. The ensemble-averaged data were collected over $2*10^8$ MC cycles.

The asymmetry of the coil shapes implies that some specific monomers prefer to move ahead of the others. We therefore made statistics on the distribution of monomer sequences locating at the maximum Y-coordinates. The results are shown in Fig. 3. One can clearly see that without driving forces two chain ends dominate the distribution, but when the deformation happens the chain middles become dominant as well. The dominant chain ends at the forefront of the coil are normal because of their least restriction along the chain and thus the highest mobility. In the force field, the significant acceleration in the middle of the chain can be attributed to an integrated interaction between two monomers consecutively connected along the chain, which must be responsible for the presently observed coil deformation.

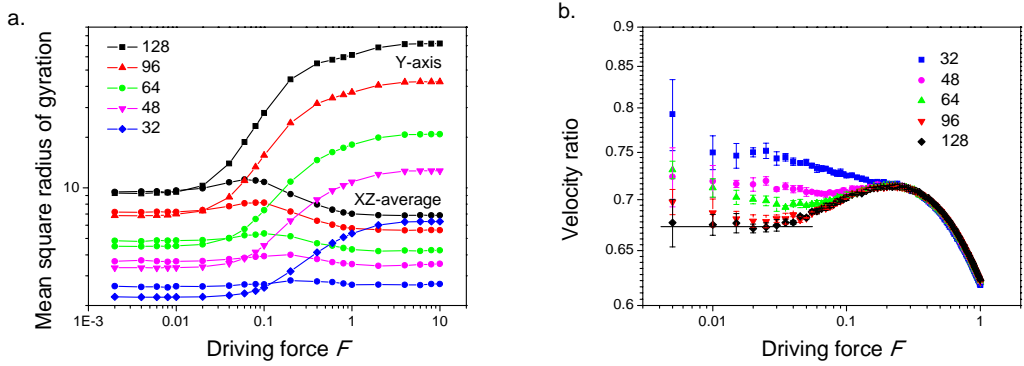

**Figure 4** **(a)** Parallel fraction (squares) and perpendicular fraction (spheres, averaged over two dimensions) of mean square radius of gyration for the single chain containing a certain amount of monomers ($N$=32, 48, 64, 96, and 128) under various driving forces. The data under each condition were collected as an ensemble average over $10^8$ MC cycles. The segments connecting the same group of symbols are drawn to guide the eyes. **(b)** Ratios of the activated velocities to the monomer expectation 1-exp(-$F$) versus the driving forces for the single chains containing different amounts of monomers (chain lengths $N$=32, 48, 64, 96, and 128) under various driving forces. The data under each condition were collected as an average over five individual observations in each step period of $10^7$ MC cycles. The straight horizontal line is drawn to guide the eyes for the subsequent shear-thinning.

In principle, more monomers in a longer chain hold higher degrees of translational freedom, which will bring a more significant effect to the diffusivity and the deformation of the chain. We furthermore observed the deformation with various chain lengths. Figure 4a demonstrates that the longer chains indeed intend to initiate deformation at weaker driving forces.

In our present simulations, the Metropolis sampling generates the expected velocity 1-exp(-$F$) of each monomer, which is the net result of moving-forward probability (one) and moving-back probability (exp(-$F$)). The expected velocity of each monomer reflects the strength of driving force. The activated velocity of the mass center of the single chain integrates over all the contributions of monomers, but appears smaller than the expected velocity of each monomer due to chain connection of the latter. The Newtonian fluid of driven polymers features with the activated shear rates proportional to the enhancement of shear stresses, resulting in constant shear viscosities as defined by their ratios. Here the shear stress corresponds to the driving force, or more specifically, the expected monomer velocity, while the shear rate corresponds to the activated velocity of coil mass center. So the ratios of the activated velocity of coil mass center to the expected velocity of each monomer exhibits a horizontal curve for the Newtonian-fluid behavior, and any deviation from the horizontal reflects some sorts of non-Newtonian-fluid behaviors.

Figure 4b summarizes the velocity ratios versus driving forces for various chain lengths. In comparison to Fig. 4a, one can see that the horizontals indicate the Newtonian-fluid behaviors of linear polymers before the occurrence of deformation. The horizontal levels decrease with the increase of chain lengths due to the extended monomer restrictions along the chain. For various chain lengths at higher driving forces, the velocity ratios sequentially merge into a master curve. The master curve first raises up and then goes down, corresponding to shear thinning (a larger velocity ratio corresponds to a lower viscosity) followed with shear thickening. Short chains (for instance, $N$=32 monomer units) jump directly from Newtonian-fluid behaviors to shear thickening without experiencing any shear-thinning. Longer chains exhibit more significant shear-thinning behaviors at weaker onset driving forces, which can be attributed to more consecutive alignments of bond orientations for a higher efficiency of sliding diffusion, similar to the consideration employed in the tube model of the Doi-Edwards theory for the interpretation of shear-thinning phenomenon of bulk polymers.[13] The chain-length-independence of shear-thickening behaviors can be understood as well by the Larson's kink-dynamics model for polymers under large extension.[28] Here surprisingly, such a simple model of driven polymers can perform coil deformation that holds those essential features considered by two theories above and reproduces well the basic non-Newtonian-fluid behaviors of flowing chain-like polymers. Apparently, the coil deformation triggers these non-linear viscoelastic behaviors, and its origin is worthy of further investigation.

In our dynamic MC simulations, monomers are randomly selected on the polymer chain to perform micro-relaxation, and each selection makes a monomer to move earlier than its neighbors on the chain, which will bring an additional acceleration effect due to their chain connection. Such an operation reflects non-synchronous activation upon Brownian motions of monomers. Without any external force field, this additional acceleration will be neutralized upon an integration of random Brownian motions of monomers, which causes no coil-deformation. Since this acceleration effect appears in principle asymmetric between chain ends and chain middles, the asymmetry of the effect will be accumulated on the flow direction in a strong-enough force field, and thus an internal tension between chain ends and chain middles will raise coil-deformation, resulting in the phenomenon observed in Fig. 3.

On the basis of linear Langevin equations of monomers, Brownian Dynamics (BD) simulations assume synchronous activation among the Brownian motions of all the monomers, which update all the bead positions at the same time. The synchronous updating of monomer positions appears as a linear integration of those linear Langevin equations of monomers, resulting in a linear Langevin equation of polymer chains. Any non-synchronous activation of Brownian motions of monomers will bring non-linear integration of monomers to the polymer chain. So far, the non-linear Langevin equations of polymer chains to describe the non-synchronous integration of linear Langenvin equations of monomer motions have not yet been established; therefore, as a preliminary investigation, BD simulations can help us to understand such an effect of non-synchronous activation. In the following, we perform BD simulations of a single phantom free-draining bead-spring polymer in a constant force field, and compare three methods of updating bead positions, i.e. synchronous updating (conventional BD simulations), random updating (like in MC simulations, the only information delivered from MC to BD simulations but no more) and sequential updating (an extreme non-synchronous case) on the driven single polymer. We found that introducing any non-synchronous updating of monomer positions into the Brownian motions of monomers will indeed cause coil-deformation of driven polymers.

## IV. TECHNIQUES OF BROWNIAN DYNAMIC SIMULATIONS

### A. Governing equations

We followed the common protocols of BD simulations of polymer chains[29-32] except for the synchronous updating of monomer positions. In details, we used the simple Rouse-chain model of polymer dynamics. The single polymer chain was treated as a string of $N$ beads consecutively connected by $N$-1 springs. All the forces exerted onto each bead, and the spring was not subjected to forces. The spring was illusory, which could pass through each other, and only the spring directly connected to the bead had an impact onto the bead. The move of the whole chain was caused by the cooperative motion of the beads, while the move of each bead followed the linear Langevin equation[29]

$$F_{drag\ i} + F_{spring\ i} + F_{Brownian\ i} + F_{driving\ i} = 0 \qquad (4)$$

Here, $F_{drag\ i}$ was the drag force, $F_{spring\ i}$ was the total spring force on the $i$-th bead, $F_{brownian\ i}$ was the random force due to thermal fluctuations of solvent molecules, $F_{driving\ i}$ was the external force exerted onto each bead. The drag force on the $i$-th bead could be given as

$$F_{drag\ i} = -\xi(r_i' - v) \qquad (5)$$

where $\xi$ was the drag coefficient of the bead, $r_i'$ was the time derivative of the position vector of the bead, $v$ was the velocity of the solvent. Here, we assumed $v = 0$.

The spring force on the $i$-th bead could be given as

$$F_{spring\ i} = f_{spring\ i} - f_{spring\ i-1} \qquad (6)$$

where the spring forces $f_{spring\ i}$ depended on the spring law. There are different spring laws. In this work, we used the conventional finitely-extensible nonlinear elastic (FENE) spring model, the law of which according to the Warner spring law was given as[30]

$$f_{sping\ i} = \frac{3k_B T}{b^2} \cdot \frac{R_i}{1 - \lambda_i^2} \qquad (7)$$

where $b$ was the root-mean-square spring length at equilibrium, $\lambda_i$ was the extension ratio expressed as

$$\lambda_i = \frac{R_i}{N_{k,s} b_k} \qquad (8)$$

Here, $b_k$ was the Kuhn length of the free-jointed chain, $N_{k,s}$ was the number of Kuhn lengths per spring, and $R_i$ was the vector of the spring, which was given as

$$R_i = |r_{i+1} - r_i| \qquad (9)$$

where $r_i$ was the position of the $i$-th bead.

The Brownian force in the simulation could be given as[31]

$$F_{Brownian\ i} = g_i \sqrt{\frac{6\xi k_B T}{\Delta t}} \qquad (10)$$

where $g_i$ was a Gaussian random vector with each component within ($-\infty$, $+\infty$), satisfying the Gaussian normal distribution with zero mean and unit variance, and $\Delta t$ was the time step.

The driving force made by the external force field was expressed similar to Eq. (1), as given by

$$F_{driving\ i} = F \cdot \delta r_i / b \qquad (11)$$

where $\delta r_i$ represents the position shift of the $i$-th bead in each time step along the direction of driving force, and the unit of driving forces $F$ was reduced in dimension by the characteristic force $f$.

**B. Simulation algorithm and parameters**

In order to simplify our BD simulations, we used the dimensionless forms of the length, the time and the force. Here, we followed the steps suggested by Panwar and Kumar.[31] The characteristic length was chosen as the equilibrium length of spring $b$. The characteristic time was

$$\tau = \frac{\xi b^2}{k_B T} \qquad (12)$$

The characteristic force was

$$f = \frac{k_B T}{b} \qquad (13)$$

The Langevin equation in dimensionless form was written as

$$\frac{dp_i}{dt} = g_i \sqrt{\frac{6}{\Delta t}} + \frac{3}{1-\lambda_i^2}(p_{i+1} - p_i) - \frac{3}{1-\lambda_{i-1}^2}(p_i - p_{i-1}) + F_{driving\ i}{}' \qquad (14)$$

where $p_i$ was the dimensionless bead position, $F_{driving\ i}{}'$ was the dimensionless driving force.

The Langevin equation decides the time evolution of the system, and we need to use an appropriate time marching method to record it. We used the first-order explicit Euler scheme. The new position of the bead was obtained from settling the following equation[31]

$$p_i(t_{k+1}) = p_i(t_k) + dp_i \qquad (15)$$

where $p_i(t_k)$ was the initial position of the $i$-th bead, $dp_i$ was the movement of the $i$-th bead during the time step, and $p_i(t_{k+1})$ was the new position after one time step.

In our BD simulations, the time step was chosen as 0.001, which had been proven to be effective.[32] The bead number was 20 and one spring in the simulation contained 100 Kuhn lengths.

**C. Preparation of the initial state**

The biased diffusion of the chain in the BD simulations began from a random coil. The random coil was obtained by the relaxation of an extended chain. The relaxation allowed all beads to perform Brownian motions under random forces, spring forces and friction forces. The mean square radius of gyration was obtained according to Eq. (2). When three orthogonal fractions of mean square radius of gyration were stabilized, the chain reached random coil. After such a relaxation process, the positions of beads were saved as for the initial state of a driven polymer.

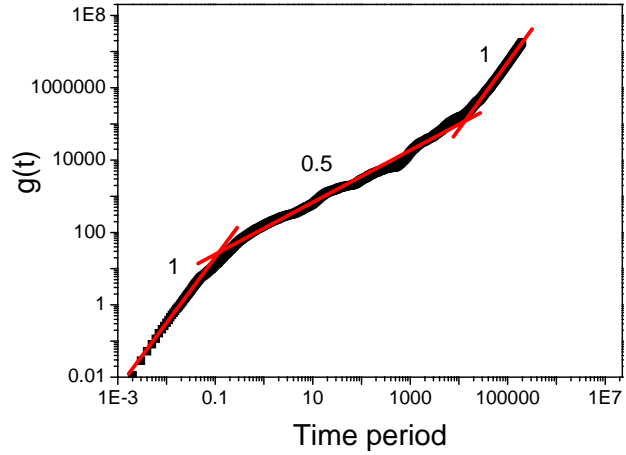

**Figure 5** Log-log plots of mean square displacements of beads for a single phantom chain containing 20 beads versus time period without any external driving force. Every data point was averaged over 100 times. The bead positions were randomly updated.

To identify the reasonably well Brownian motions of the short chain in our BD simulations with a random updating scheme, we calculated the mean square displacement of monomer $g(t)$ according to Eq. (3) under zero driving force. As shown in Fig. 5, the curve changes its slopes from one to 0.5 and then back to one, following the exact time-scaling laws of the Rouse-chain model.

**V. RESULTS OF BROWNIAN DYNAMICS SIMULATIONS**

The deformation of the single coil can be traced by the Y-axis fractions of mean square radius of gyration with the enhancement of driving forces along Y-axis, as shown in Fig. 6a. Among three sampling methods to update bead positions, the sequential-updating method as an extreme non-synchronous case raises the most significant deformation under strong driving forces, and the random-updating method as similar with MC simulations brings also deformation to the coil, although its deformation is far from saturated; in contrast, the conventional synchronous-updating method keeps the coil shape unchanged under strong driving forces. Again, the deformation appears steady and reversible on the way back with the weakening of driving forces.

Figure 6b provides a visual inspection on the spatial distributions of monomers around the mass center of the chain under strong driving forces. Using the non-synchronous sampling methods like random updating and sequential updating, both polymer coils appear as consistent with the results of MC simulations, while the polymer coil by using the synchronous-updating method remains in a circle profile, indicating no deformation.

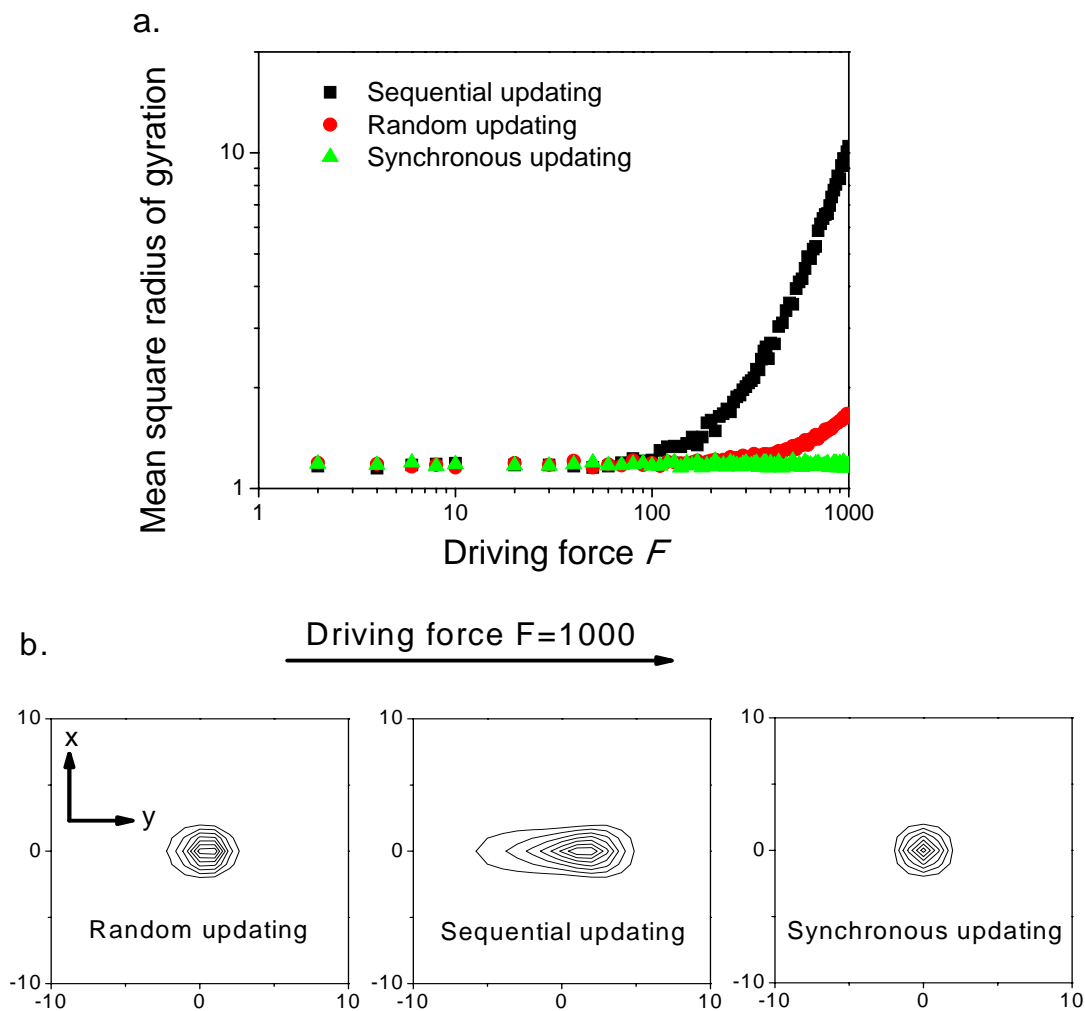

**Figure 6 (a)** Y-axis fractions of mean square radius of gyration of a single chain containing 20 beads under various strengths of driving forces with three updating methods labeled by different symbols. The relaxation time of the chain under each driving force was 200,000. Each data point was collected as an ensemble average over 200,000 times. **(b)** Contour plots of two-dimensional projections of bead densities at XY plane around the mass center of a single phantom chain containing 20 beads under strong driving forces (here 1000 units). Three kinds of sampling methods were used as denoted in the figures: random updating, sequential updating, and synchronous updating. The data were collected as an ensemble average over $2*10^7$ times. The driving forces were applied along Y-axis.

We further observed the probability of the bead sequences at the forefront of Y-axis upon biased diffusion, in comparison to the corresponding cases without biased. The results in Fig. 7 show that in the sequential-updating method the last sequenced bead moves the fastest, while in the random-updating method the middle bead shows the highest probability to move forward. In the synchronous-updating method, the probability distribution is not changed by employing driving forces.

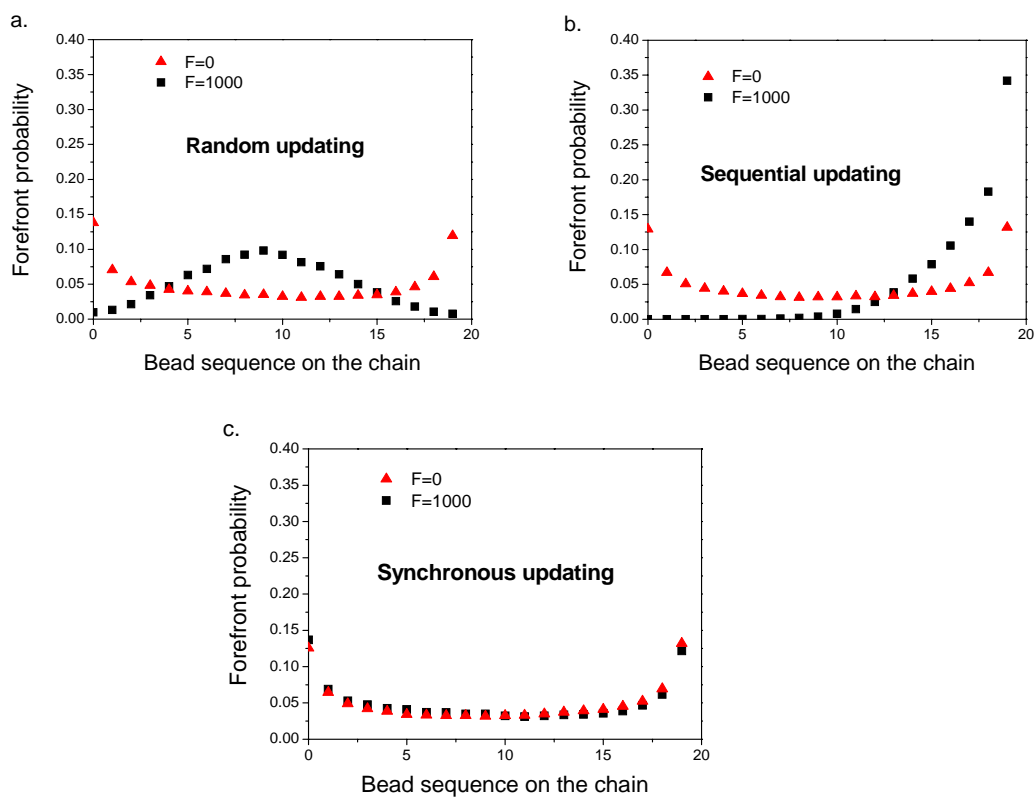

**Figure 7** Distributions of the forefront probability of each bead on the sequences of a single phantom chain containing 20 beads with **(a)** random updating, **(b)** sequential updating, and **(c)** synchronous updating. Different driving forces ($F$=0 or 1000 units as labeled) were employed on Y-axis.

In the non-synchronous-updating methods, the reason for the change of forefront beads could be that the biased-jumping bead accelerates its neighbors jumping later. As an extreme case in the sequential-updating method, due to the bead position updating from the head to the tail, the tail attains the largest acceleration effect accumulated along the chain, thus it shows the highest probability of moving forward, similar like cracking the whip. In the more realistic random-updating method, this synergetic effect can only be accumulated in the middle of the chain, similar with what we have observed in MC simulations above. Therefore, non-synchronous updating of bead positions appears as responsible for the deformation of the coil.

The non-synchronous activation is an arbitrary operation added onto the conventional BD simulations, which demonstrates the source of a non-linear effect in both MC and MD dynamics. Understanding the implication of this source in the complicated real situation will bring us new insights into non-linear dynamics of real polymer systems. Without strong driving forces, the "cracking-the-whip" effects are neutralized upon random Brownian motions, and thus the synchronous and non-synchronous BD simulations are consistent with each other. Under strong driving forces, the "cracking-the-whip" acceleration effects will be integrated along a specific direction, causing a non-linear dynamics deviating from the linear dynamics. Therefore, an additional non-linear effect has been brought into the integration of Langevin equations of BD simulations upon non-synchronous updating.

The random updating of non-synchronous activation of monomers corresponds to random perturbations in the environmentally dependence free energy landscape or mobility among the Brownian motions of monomers. In the real bulk polymer phase, any instant spatial distribution of monomers appears always inhomogeneous, so the local density environments can be different among the monomers belonging to the same chain, raising heterogeneous dynamics. In addition, thermal/density fluctuations in the bulk polymer phase generally enhance the heterogeneous dynamics within each polymer. The heterogeneous dynamics has been widely investigated by both theory and experiments in the scenario of supercooled liquids.[33] In heterogeneous dynamics, those more biased monomers will bring an acceleration to their neighbors on the chain, and an integration of this "cracking-the-whip" effect along the direction of driving forces raises an internal tension between chain ends and chain middles for the observed coil-deformation. Therefore, our observation holds a fundamentally important meaning to the non-linear dynamics of real driven polymers.

## VI. CONCLUSION

By means of a combinational investigation of dynamic MC simulations and BD simulations, we assigned the mechanism of force-induced polymer deformation to the intrinsic internal tension between chain ends and chain middles raised by the "cracking-the-whip" effect upon random non-synchronous activation among driven monomers. The deformation appears reversible under variable driving forces, and the onset strengths of driving forces decay with the increase of chain lengths. The thinning-followed-with-thickening characters of non-Newtonian-fluid behaviors of polymer flows can be reproduced by this simple model. This result implies that the related non-linear viscoelasticity of polymer fluids could be originated from the coil deformation raised by this "cracking-the-whip" effect. In association to the more complicated reality, the non-synchronous activations of monomers in our simple ideal model correspond to the generally existed heterogeneous dynamics within each polymer chain. In this sense, our approach paves a new way to understand the profound non-linear viscoelastic behaviors of driven polymers.

## ACKNOWLEDGMENTS

The financial support of National Natural Science Foundation of China (Grant No. 20825415 and 21274061) and the National Basic Research Program of China (Grant No. 2011CB606100) is appreciated.